\documentstyle[12pt]{article}

\newcommand {\e} {\mbox{\rm e}}

\newcounter{eq}
\newcounter{sc}

\def\overleftrightarrow#1{\vbox{\ialign{##\crcr
 $\leftrightarrow$\crcr\noalign{\kern-1pt\nointerlineskip}
 $\hfil\displaystyle{#1}\hfil$\crcr}}}

\newlength{\minitwocolumn}
\begin{document}

\begin{flushright}
DPUR/TH/61\\
October, 2018\\
\end{flushright}
\vspace{20pt}

\pagestyle{empty}
\baselineskip15pt

\begin{center}
{\large\bf  Higgs Potential from Wick Rotation in Conformal BSM
\vskip 1mm }

\vspace{20mm}

Ichiro Oda\footnote{
           E-mail address:\ ioda@sci.u-ryukyu.ac.jp
                  }

\vspace{10mm}
           Department of Physics, Faculty of Science, University of the 
           Ryukyus,\\
           Nishihara, Okinawa 903-0213, Japan\\

\end{center}


\vspace{10mm}
\begin{abstract}

It is well known that in order to make the path integral of general relativity converge,
one has to perform the Wick rotation over the conformal factor in addition to the more
familiar Wick rotation of the time axis to pass to the space-time with Euclidean signature.
In this article, we will apply this technique to a scalar field in the conformally invariant scalar-tensor
gravity with a conformally invariant beyond-standard-model (BSM). It is then shown that
a potential term in the conformally invariant potential, which corresponds to the Higgs mass term 
in the Higgs potential of the standard model (SM), can have a negative coefficient. 
The change of sign of the potential term naturally induces spontaneous symmetry breakdown 
of the electroweak gauge symmetry after symmetry breaking of conformal symmetry (local scale symmetry) 
via the Coleman-Weinberg mechanism around the Planck scale.  The present study might shed light on the
fact that the existence of a stable vacuum in quantum gravity is relevant to that in the SM.

\end{abstract}

\newpage
\pagestyle{plain}
\pagenumbering{arabic}


The discovery \cite{Atlas, CMS} of a relatively light Higgs particle with properties consistent with 
the standard model (SM) has marked a significant milestone in the history of particle physics.  
The SM of particle physics, which describes the electro-magnetic, weak and strong interactions in a
concise manner, has passed a series of stringent tests so far. It is remarkable that as the parameters
in the model have been measured precisely by many of experiments, points of disagreement existed in the past
have completely faded away and the SM has been put on the more sound ground. 
  
From the viewpoint of the SM, it appears that the Planck scale $M_{Pl}$ is a special point in the sense 
that\footnote{With the top mass, $m_t = 173 GeV$, the renormalization group equation for the Higgs
self-coupling constant $\lambda$ implies that $\lambda$ becomes negative around $10^{11} GeV$ \cite{Buttazzo},
whereas with the current uncertainty of experiments, the top quark might have the lighter value, $m_t =
170 GeV$, and then $\lambda$ becomes zero around the Planck mass scale.}
\begin{enumerate}
  \item Scalar self-coupling is zero: $\lambda(M_{Pl}) = 0$,
  \item Its beta function is zero: $\frac{d \lambda}{dt} |_{M_{Pl}} \equiv \beta(M_{Pl}) = 0$,
  \item Higgs bare mass is zero \cite{Hamada}: $m^2(M_{Pl}) = 0$.
\end{enumerate}  
These facts suggest that the SM might secretely know the physics at the Planck scale even if it does not involve
gravity. Here it is of interest to reverse this viewpoint and suppose that the physics at the Planck 
scale, which we call {\it{quantum gravity}}, might give us some useful information on the SM or 
the construction of a theory beyond the standard model (BSM).  Under such a situation, there might be
some aspects of the SM that do not involve gravity directly but nevertheless require some information from 
quantum gravity. As such an example, in this article, we shall shed light on the sign problem of the tachyonic mass
term in the Higgs potential of the SM.

The above observations also suggest that it would be conceivable that the SM is the low-energy limit of a distinct 
special theory with a global scale symmetry at the Planck scale. However, as stressed in our previous work \cite{Oda0},
both no-hair theorem of quantum black holes and the fact that in string theory any additive global symmetries
are either gauge symmetries or explicitly violated in a tacit way seem to insist that a global scale symmetry
must be promoted to a local scale symmetry, which we call {\it{conformal symmetry}} in this article.\footnote{We 
have already constructed such models with scale symmetries at the classical level \cite{Oda1}-\cite{Oda4}.}  

As far as experiments based on accelerators are concerned, the SM does a rather excellent
job of accounting for various kinds of particle phenomena. The objection to the opinion that the SM is a complete 
theory mainly comes from a theoretical side. In particular, the SM has a number of arbitrary parameters which cannot be
explained theoretically but are fixed only by measurements. For instance, the renormalizability of the SM
requires that the Higgs potential takes the simple form 
\begin{eqnarray}
V(H) = m^2 (H^\dagger H) + \frac{\lambda_H}{2} (H^\dagger H)^2,
\label{Higgs potential}
\end{eqnarray}
up to radiative corrections. For spontaneous symmetry breakdown to occur, the renormalized value of the
mass parameter $m^2$ must be negative. But the parameter $m^2$ could have either sign; there is no 
logic that we prefer one sign to the other. We should therefore answer the question why the mass parameter $m^2$
is negative in order to understand the Higgs mechanism completely \cite{Peskin}. 
  
This issue is closely related to the gauge hierarchy problem. The SM action is invariant under a global scale
transformation except the Higgs mass term, that is, one could say that our world is almost scale 
invariant.\footnote{It is easy to extend a global scale symmetry to conformal symmetry by introducing the conformally
invariant coupling between the Higgs field and the scalar curvature as seen shortly.}  Indeed, Bardeen has advocated 
the idea that instead of supersymmetry, the global scale symmetry might be a fundamental symmetry and play an important role
in the naturalness problem \cite{Bardeen}.  With scale invariance, the Higgs potential consists of solely the second term
in Eq. (\ref{Higgs potential}), so the mass correction is only the logarithmic divergence rather than quadratic one, thereby
alleviating the gauge hierarchy problem.

In our recent study \cite{Oda0},  it has been shown that both the Planck and electroweak scales can be generated from
conformal gravity via the Coleman-Weinberg mechanism \cite{Coleman}, which explicitly breaks conformal symmetry, and 
via the conformally invariant potential corrected by the Coleman-Weinberg mechanism. To this end, symmetry breakings 
must occur at two steps: at the first step, the Planck scale is generated by radiative corrections 
associated with gravitons, by which conformal symmetry is explicitly broken. At the second step, the electroweak gauge symmetry
is spontaneously broken via the conformally invariant Higgs potential modified by the Coleman-Weinberg mechanism. 
The huge hierarchy between the two scales is explained in terms of a very tiny coupling between the scalar and
Higgs fields.

As in many of similar scale-invariant BSM models \cite{Hempfling, Shaposhnikov}, a tantalizing aspect of this study is that 
the conformally invariant scalar potential does not give rise to spontaneous symmetry breakdown of the electroweak symmetry 
unless we assume that a term in the potential, which exactly corresponds to the Higgs mass term in Eq. (\ref{Higgs potential}), 
has a negative coefficient. Since we believe that this sign problem would be clarified in future, we should derive the negative 
coefficient by some mechanism within the framework of the BSM. Much of the impetus for the present work stemmed from 
the realization that the scalar field in our conformal BSM is very similar to a conformal factor of the metric perturbation 
in general relativity and both of them are non-dynamical fields at least at the classical level. Just as the Wick rotation over 
the conformal factor guarantees that the Euclidean Einstein-Hilbert action is bounded from below and consequently makes 
the path integral be convergent in general relativity  \cite{Gibbons, Mazur, Percacci}, we expect that the Wick rotation 
over the scalar field might enable us to flip the sign in front of the Higgs mass term from positive to negative, thereby making it 
possible to trigger the spontaneous symmetry breaking of the electroweak symmetry in a natural way.     
This possibility was briefly suggested in the previous work \cite{Oda0}, but was not investigated in detail. The purpose of 
the present short article is to pursue this possibility and spell out its detail and result. 
Incidentally, this procedure cannot be applied to the case of a global scale symmetry since in this case the scalar field
is in general a dynamical field even in the classical regime.  
      
Now let us start with the following conformally invariant Lagrangian density:
\begin{eqnarray}
\frac{1}{\sqrt{-g}} {\cal L}_C &=& - \frac{1}{2 \xi^2} C_{\mu\nu\rho\sigma} C^{\mu\nu\rho\sigma}
+ \frac{1}{12} \phi^2 R + \frac{1}{2} g^{\mu\nu} \partial_\mu \phi \partial_\nu \phi
- \frac{1}{6} (H^\dagger H) R - g^{\mu\nu} (D_\mu H)^\dagger (D_\nu H)
\nonumber\\
&+& V(\phi, H) + L_m.
\label{Orig-Lag}
\end{eqnarray}
Here $\xi$ is a dimensionless coupling constant, and the conformal tensor (or Weyl tensor) $C_{\mu\nu\rho\sigma}$ is defined as
\begin{eqnarray}
C_{\mu\nu\rho\sigma} = R_{\mu\nu\rho\sigma} - ( g_{\mu [\rho} R_{\sigma] \nu} 
- g_{\nu [\rho} R_{\sigma] \mu} ) + \frac{1}{3} g_{\mu [\rho} g_{\sigma] \nu} R,
\label{Conformal tensor}
\end{eqnarray}
where $A_{[\mu} B_{\nu]} = \frac{1}{2} ( A_\mu B_\nu - A_\nu B_\mu )$, $\mu, \nu, \cdots = 0, 1, 2, 3$,
$R_{\mu\nu\rho\sigma}, R_{\mu\nu}$ and $R$ are the Riemann tensor, the Ricci tensor and the scalar curvature,
respectively.\footnote{We will follow the conventions and notation by Misner et al. \cite{MTW}.}   
We have introduced two scalar fields, one of which is the Higgs doublet $H$, $D_\mu$ is a covariant derivative 
including the SM gauge fields, and $L_m$ denotes the remaining Lagrangian density of the SM but the Higgs 
mass term, which is also conformally invariant. The second and third terms on the RHS of Eq. (\ref{Orig-Lag}) 
represent the conformally invariant scalar-tensor gravity with a positive Newton constant and a scalar ghost $\phi$.
And the fourth and fifth terms correspond to the conformally invariant terms for the Higgs field.

Moreover, the new potential $V(\phi, H)$ beyond the SM, which is conformally invariant as well, is added and has the form
\begin{eqnarray}
V(\phi, H) = \frac{\lambda_\phi}{4 !} \phi^4  + \lambda_{H \phi} (H^\dagger H) \phi^2  
+ \frac{\lambda_H}{2} (H^\dagger H)^2,
\label{V(phi, H)}
\end{eqnarray}
where all the coupling constants $\lambda_i (i = \phi, H \phi, H)$ are dimensionless. Note that the requirement of 
conformal invariance, gauge invariance and renormalizability uniquely fixes the form of the potential. (The removal 
of the requirement of renormalizability would allow the presence of non-polynomial terms such as $\frac{(H^\dagger H)^3}{\phi^2}$.)
In order to obtain a stable ground state, we require that for arbitrary positive scalar fields, the potential
is positive, $V(\phi, H) > 0$. This constraint on the potential implies that all the coupling constants
must obey the relation
\begin{eqnarray}
\lambda_i > 0.
\label{Coupling}
\end{eqnarray}

Next, it is straightforward to prove that the action, $S_c \equiv \int d^4 x {\cal L}_C$, is invariant under conformal 
transformation:
\begin{eqnarray}
g_{\mu\nu} \rightarrow \Omega^2 (x) g_{\mu\nu},  \quad 
\phi \rightarrow \Omega^{-1} (x) \phi, \quad H \rightarrow \Omega^{-1} (x) H, \quad
A_\mu \rightarrow A_\mu.
\label{Conf transf}
\end{eqnarray}
Then, let us notice that the new composite metric 
\begin{eqnarray}
\hat g_{\mu\nu} \equiv \frac{1}{6 M_{Pl}^2} \phi^2 g_{\mu\nu},  
\label{Conf-inv metric}
\end{eqnarray}
is invariant under the conformal transformation (\ref{Conf transf}). The factor $\frac{1}{6 M_{Pl}^2}$ is inserted
for the dimensional alignment and later convenience. With this new metric $\hat g_{\mu\nu}$, the Lagrangian density 
for the conformally invariant scalar-tensor gravity can be rewritten as the Einstein-Hilbert form
\begin{eqnarray}
{\cal L}_{CST} &=& \sqrt{-g} \left( \frac{1}{12} \phi^2 R + \frac{1}{2} g^{\mu\nu} \partial_\mu \phi 
\partial_\nu \phi \right)
\nonumber\\
&=& \frac{1}{16 \pi G} \sqrt{- \hat g} \hat R,
\label{Con-inv ST}
\end{eqnarray}
where we have set $\frac{M_{Pl}^2}{2} \equiv \frac{1}{16 \pi G}$. To prove this equation, we have used the following properties
under the conformal transformation (\ref{Conf transf}):
\begin{eqnarray}
\sqrt{-g} \rightarrow \Omega^4 (x) \sqrt{-g},  \quad 
R \rightarrow \Omega^{-2} (x) \left( R - 6 \Omega^{-1} (x) \nabla^2 \Omega(x) \right).
\label{Conf transf 2}
\end{eqnarray}
Furthermore, because of conformal symmetry, the total classical Lagrangian density ${\cal L}_C$ in Eq. (\ref{Orig-Lag}) 
can be rewritten as the same form as before except ${\cal L}_{CST}$ when expressed in terms of the metric tensor $\hat g_{\mu\nu}$ 
\begin{eqnarray}
\frac{1}{\sqrt{- \hat g}} {\cal L}_C &=& - \frac{1}{2 \xi^2} \hat C_{\mu\nu\rho\sigma} \hat C^{\mu\nu\rho\sigma}
+ \frac{1}{16 \pi G} \hat R  - \frac{1}{6} (\hat H^\dagger \hat H) \hat R - \hat g^{\mu\nu} (D_\mu \hat H)^\dagger (D_\nu \hat H)
\nonumber\\
&+& V(\hat \phi, \hat H) + L_m,
\label{Orig-Lag 2}
\end{eqnarray}
where we have used $\hat D_\mu = D_\mu$ owing to $\hat A_\mu = A_\mu$. 

Our task now is to consider the gravitational path integral on the basis of the metric tensor $\hat g_{\mu\nu}$.
We shall perform a Wick rotation and henceforth work on the manifold with the Euclidean signature metric $(+, +, +, +)$.\footnote{Since 
the concept of time has no physical meaning in general relativity because of diffeomorphisms, defining its Euclidean continuation as 
an analytic continuation of the time coordinate is not a natural prescription. A more reasonable prescription is to regard 
the Wick rotation as an analytic continuation of not the time coordinate but metric tensor.}
With the Euclidean signature, the Lagrangian density (\ref{Orig-Lag 2}) can be cast to the form
\begin{eqnarray}
\frac{1}{\sqrt{\hat g}} {\cal L}_C^{(E)} &=& - \frac{1}{2 \xi^2} \hat C_{\mu\nu\rho\sigma} \hat C^{\mu\nu\rho\sigma}
- \frac{1}{16 \pi G} \hat R  + \frac{1}{6} (\hat H^\dagger \hat H) \hat R - \hat g^{\mu\nu} (D_\mu \hat H)^\dagger (D_\nu \hat H)
\nonumber\\
&+& V(\hat \phi, \hat H) + L_m.
\label{Eucl-Orig-Lag}
\end{eqnarray}
With respect to this Lagrangian density, it is worthwhile to emphasize that the square of conformal tensor is certainly positive definite so that 
the action of conformal gravity is bounded from below. However, this positiveness of the conformal gravity action does not mean
the positiveness of the total Lagrangian density (\ref{Eucl-Orig-Lag}) since the conformal gravity does not include the conformal factor
due to conformal invariance.  To put differently, the total action corresponding to the Lagrangian density (\ref{Eucl-Orig-Lag}) is not bounded 
from below owing to the presence of the conformal factor which is involved only in the Einstein-Hilbert term. 
Thus, even in the present situation we have to rely on the Wick rotation over the conformal factor to make the path integral converge
as in general relativity.  It is known that the Euclidean action for matter fields is positive semi-definite so it is free
from the conformal factor problem \cite{Gibbons}, which will be explained later. Since we are interested in the conformal factor problem,
we shall henceforth pay attention to only the Einstein-Hilbert action and ignore the other actions.

At a more elementary level, we wish to define the gravitational path integral by the perturbation theory. To do that,
we make use of the background field method and therefore split as
\begin{eqnarray}
g_{\mu\nu} = \bar g_{\mu\nu} + h_{\mu\nu},  \quad 
\phi = \bar \phi + \varphi,
\label{Background method}
\end{eqnarray}
where we have defined $\bar \phi = \sqrt{6} M_{Pl}$.  Then, the conformally invariant new metric takes the form
\begin{eqnarray}
\hat g_{\mu\nu} = \bar g_{\mu\nu} + \hat h_{\mu\nu} + {\cal{O}}(\hat h^2),
\label{Background new metric}
\end{eqnarray}
with $\hat h_{\mu\nu}$ being defined as 
\begin{eqnarray}
\hat h_{\mu\nu} = h_{\mu\nu} + \sqrt{\frac{2}{3}} \frac{1}{M_{Pl}} \bar g_{\mu\nu} \varphi.
\label{hat h}
\end{eqnarray}
In this article, we shall work with a perturbation theory where it is assumed that
\begin{eqnarray}
| h_{\mu\nu} | \ll 1, \quad \varphi \ll M_{Pl}, \quad |\hat h_{\mu\nu}| \ll 1.
\label{Perturb}
\end{eqnarray}

In a curved space-time, it is more convenient to introduce the York decomposition \cite{York, Percacci}:
\begin{eqnarray}
\hat h_{\mu\nu} = \hat h_{\mu\nu}^{TT} + \bar \nabla_\mu \xi_\nu + \bar \nabla_\nu \xi_\mu
+ \bar \nabla_\mu \bar \nabla_\nu \sigma - \frac{1}{4} \bar g_{\mu\nu} \bar \nabla^2 \sigma
+ \frac{1}{4} \bar g_{\mu\nu} \hat h.
\label{York}
\end{eqnarray}
Here, our conventions are the following: all indices are raised and lowered with the background metrics 
$\bar g^{\mu\nu}$ and $\bar g_{\mu\nu}$, and the trace is defined as $\hat h \equiv \bar g^{\mu\nu} 
\hat h_{\mu\nu}$. Moreover, $\bar \nabla_\mu$ denotes the Levi-Civita connection of the background metric 
$\bar g_{\mu\nu}$, $\hat h_{\mu\nu}^{TT}$ is transverse and traceless, and $\xi_\mu$ is transverse:
\begin{eqnarray}
\bar \nabla^\mu \hat h_{\mu\nu}^{TT} = 0, \quad \bar g^{\mu\nu} \hat h_{\mu\nu}^{TT} = 0,
\quad \bar \nabla^\mu \xi_\mu = 0.
\label{TT-condition}
\end{eqnarray}
Given the York decomposition (\ref{York}), let us define the conformal factor in the metric $\hat g_{\mu\nu}$ by
\begin{eqnarray}
\hat s \equiv \hat h - \bar \nabla^2 \sigma
= h - \bar \nabla^2 \sigma + 4 \sqrt{\frac{2}{3}} \frac{1}{M_{Pl}} \varphi.
\label{Conf-factor}
\end{eqnarray}
For simplicity, in what follows, we will assume the background metric $\bar g_{\mu\nu}$ to belong to the Einstein 
space.\footnote{In order to consider the Einstein space and make the argument clear, we add the cosmological term to
the action. It is easy to see that this modification does not change the final result.} The Einstein space, which is defined 
as the space satisfying the Einstein equation with a cosmological constant $\Lambda$   
\begin{eqnarray}
\bar R_{\mu\nu} = \Lambda \bar g_{\mu\nu}, 
\label{Eins-eq}
\end{eqnarray}
is a classical solution to the field equation of the Einstein-Hilbert action.
   
Under these conditions, the quadratic part of the Euclidean Einstein-Hilbert action can be evaluated to be
\cite{Mazur, Percacci}
\begin{eqnarray}
S^{(2)}_{EH} (\hat h; \bar g) = \frac{1}{32 \pi G} \int d^4 x \sqrt{\bar g} \left[ \frac{1}{2} \hat h_{\mu\nu}^{TT}
( \bar \Delta_{L_2} - 2 \Lambda ) \hat h^{TT \mu\nu} - \frac{3}{16} \hat s ( - \bar \nabla^2 - \frac{4}{3}
\Lambda ) \hat s \right],
\label{Quad-EH}
\end{eqnarray}
where the operator $\Delta_{L_2}$ represents the Lichnerowicz Laplacian acting on generic second-rank symmetric tensors
$T_{\mu\nu}$ which is concretely defined as
\begin{eqnarray}
( \bar \Delta_{L_2} T)_{\mu\nu} = - \bar \nabla^2 T_{\mu\nu} + \bar R_\mu \, ^\rho T_{\rho\nu} 
+ \bar R_\nu \, ^\rho T_{\rho\mu} - 2 \bar R_\mu \, ^\rho \, _\nu \, ^\sigma T_{\rho\sigma}.
\label{Lichnerowicz}
\end{eqnarray}
Now it is clear what the conformal factor problem in general relativity is. The first term in Eq.  (\ref{Quad-EH})
is positive whereas the second one is not so. A negative kinetic term in Euclidean signature usually means 
negative energy in Lorentzian signature, but there is no such pathology in general relativity. This issue
is not restricted to perturbation theory and has a root that the full Euclidean Einstein-Hilbert action
is unbounded from below \cite{Gibbons}. Actually, under the conformal transformation (\ref{Conf transf}),
the integrand of the Einstein-Hilbert action is transformed as
\begin{eqnarray}
\sqrt{-g} R \rightarrow \sqrt{-g} \Omega^2 (x) \left( R - 6 \Omega^{-1} (x) \nabla^2 \Omega(x) \right),
\label{Conf transf 3}
\end{eqnarray}
where we have used Eq. (\ref{Conf transf 2}). One sees that this quantity can be as negative as one wants by
selecting a rapidly varying conformal factor $\Omega(x)$. 

There could be several resolutions to this problem. A well-known resolution is that the integration over 
the conformal factor is rotated in the complex plane in order to make the integrand converge \cite{Gibbons}.
A more detailed analysis has been done at least at the one-loop level in perturbation theory in \cite {Mazur},
which we shall follow in this paper. At first sight, the conformal factor $\hat s$, which is a scalar field invariant 
under diffeomorphisms, is a dynamical, propagating mode, but it is an illusion. Indeed, the Jacobian associated with the change
of variables turns out to include a determinant \cite{Mazur, Percacci} 
\begin{eqnarray}
\sqrt{\frac{ \det( - \bar \nabla^2 - \frac{4}{3} \Lambda )}{\det ( - \bar \nabla^2)} },
\label{Det}
\end{eqnarray}
whose numerator precisely cancels the determinant coming from the Gaussian integration over the conformal
factor $\hat s$ in Eq. (\ref{Quad-EH}). Thus, the Wick rotation over $\hat s$ is then justified at least at one-loop order 
due to the fact that $\hat s$ is not a dynamical field \cite{Mazur}. 

To proceed further, let us notice that as seen in Eq.  (\ref{Conf-factor}), the Wick rotation over 
the conformal factor $\hat s$ in the conformally invariant metric $\hat g_{\mu\nu}$ means the simultaneous Wick rotation over 
the conformal factor in the original metric $g_{\mu\nu}$, $s \equiv h - \bar \nabla^2 \sigma$, and the scalar field $\varphi$:
\begin{eqnarray}
\hat s \rightarrow i \hat s \Longleftrightarrow s \rightarrow i s,  \quad \varphi \rightarrow i \varphi.
\label{Wick-rot}
\end{eqnarray}
Since, with the Euclidean signature metric, the conformally invariant scalar-tensor gravity (\ref{Con-inv ST}) reads
\begin{eqnarray}
{\cal L}_{CST}^{(E)} &=& \sqrt{g} \left( - \frac{1}{12} \phi^2 R + \frac{1}{2} g^{\mu\nu} \partial_\mu \phi 
\partial_\nu \phi \right)
\nonumber\\
&=& - \frac{1}{16 \pi G} \sqrt{\hat g} \hat R,
\label{Euc-Con-inv ST}
\end{eqnarray}
performing the Wick rotation over $\varphi$ and using Eq. (\ref{Background method}) lead to the expression
\begin{eqnarray}
{\cal L}_{CST}^{(E)} &=& \sqrt{g} \left[ + \frac{1}{12} ( \varphi - i \bar \phi )^2 R 
- \frac{1}{2} g^{\mu\nu} \partial_\mu \varphi \partial_\nu \varphi \right]
\nonumber\\
&=& \sqrt{g} \left[ + \frac{1}{12} \varphi^2 R - \frac{1}{2} g^{\mu\nu} \partial_\mu \varphi \partial_\nu \varphi \right],
\label{Wick-Euc-Con-inv ST}
\end{eqnarray}
where at the last step we have performed the shift of variables, $\varphi \rightarrow \varphi + i \bar \phi$, 
in the functional measure ${\cal{D}} \varphi$ of the path integral.\footnote{Of course, it is possible to perform this shift 
of variables since the integrand in the path integral is in general complex with the analytic continuation. Since the present 
integral is of the Gaussian form with respect to $\varphi$, a suggestive formula would be $1 = \sqrt{\frac{a}{\pi}} 
\int_{-\infty}^{+\infty} d x \, \e^{- a x^2} = \sqrt{\frac{a}{\pi}} \int_{-\infty}^{+\infty} d x \, \e^{- a (x - i b)^2}$ 
where $a, b$ are some real numbers.}  
It is more convenient to go back to Lorentzian signature
\begin{eqnarray}
{\cal L}_{CST} = - \sqrt{- g} \left[ \frac{1}{12} \varphi^2 R + \frac{1}{2} g^{\mu\nu} \partial_\mu \varphi 
\partial_\nu \varphi \right],
\label{Wick-Lor-Con-inv ST}
\end{eqnarray}
which implies that compared with Eq. (\ref{Con-inv ST}), we have now obtained the conformally
invariant scalar-tensor gravity with a negative Newton constant and a normal scalar field.  
However, this is also an illusion. Actually it is strange that via the Wick rotation over the scalar field
we can obtain the conformally invariant scalar-tensor gravity with the opposite properties since
the Wick rotation does not change the physical contents of the theory at all. The source of misunderstanding
can be found in Eq.  (\ref{Conf-inv metric}), which becomes after the Wick rotation and the shift of variables
\begin{eqnarray}
\hat g_{\mu\nu} = - \frac{1}{6 M_{Pl}^2} \varphi^2 g_{\mu\nu}.
\label{Conf-inv metric 2}
\end{eqnarray}
This relation shows that the overall sign in the metric has been changed, which is not allowed in the
conformal transformation (\ref{Conf transf}).
The proper procedure is first to make a transformation 
$g_{\mu\nu} \rightarrow - g_{\mu\nu}$ and then perform the conformal transformation to reach (\ref{Conf-inv metric 2}).
It is easy to see that the transformation $g_{\mu\nu} \rightarrow - g_{\mu\nu}$ changes Eq.  (\ref{Wick-Lor-Con-inv ST}) 
to be the form
\begin{eqnarray}
{\cal L}_{CST} = \sqrt{- g} \left[ \frac{1}{12} \varphi^2 R + \frac{1}{2} g^{\mu\nu} \partial_\mu \varphi 
\partial_\nu \varphi \right],
\label{Wick-Lor-Con-inv ST 2}
\end{eqnarray}
which is nothing but the conformally invariant scalar-tensor gravity with a positive Newton constant and a scalar ghost
as in (\ref{Con-inv ST}). Incidentally, as seen shortly, with the symmetry breaking of conformal symmetry, 
$\varphi = \langle \varphi \rangle$, the first term in Eq.  (\ref{Wick-Lor-Con-inv ST 2}) yields the conventional Einstein-Hilbert term 
with the positive Newton constant, which is also bounded from below by the Wick rotation $s \rightarrow i s$ in Eq.  (\ref{Wick-rot}) 
in case of the Euclidean signature metric. 

The most appealing point in this article is that as the result of the above Wick rotation and the shift of variables over $\varphi$,
the potential term beyond the SM, Eq. (\ref{V(phi, H)}), can take the form
\begin{eqnarray}
V(\varphi, H) = \frac{\lambda_\phi}{4 !} \varphi^4  - \lambda_{H \phi} (H^\dagger H) \varphi^2  
+ \frac{\lambda_H}{2} (H^\dagger H)^2,
\label{V(varphi, H)}
\end{eqnarray}
where the sign in front of the second term on the RHS has flipped from positive to negative,
which would provide us with a natural symmetry breaking mechanism for the electroweak gauge symmetry. It is of interest
that the gravitational physics provides the important effect for the generalized Higgs potential of the BSM.
 
At this stage, it is worth reviewing our previous work \cite{Oda0} where it was shown that both the Planck and
electroweak mass scales can be generated by starting with the present formulation of the BSM. We can envision the process 
of symmetry breaking as two independent steps. At the first step, around the Planck scale, conformal symmetry is explicitly 
broken via the Coleman-Weinberg mechanism, thereby generating the Planck scale and general relativity. Next, at the second step,
around the electroweak scale, the electroweak gauge symmetry is spontaneously broken by the potential in Eq. (\ref{V(phi, H)}) 
which is modified by the Coleman-Weinberg mechanism at the first step of the symmetry breaking.  The key observation is that 
at the second step of the symmetry breaking, in order to trigger spontaneous symmetry breakdown of the electroweak symmetry, 
it was necessary to replace $\lambda_{H \phi}$ with $- \lambda_{H \phi}$ in Eq. (\ref{V(phi, H)}) in an ado hoc manner, 
which is assumed in many of scale-invariant models as well \cite{Hempfling, Shaposhnikov}. 
By contrast, in the formulation at hand, the ado hoc replacement is naturally derived in terms of the Wick rotation over the scalar field 
$\varphi$ which is allowed since the scalar field is a non-dynamical field \cite{Mazur}.  This fact is also understood from the fact that
the scalar field $\phi$ in the conformally invariant scalar-tensor gravity is a gauge freedom associated with conformal symmetry. 
This situation is very similar to that of general relativity in the sense that the conformal factor is a gauge freedom associated 
with diffeomorphisms.

Now we are ready to present the second step of the symmetry breaking of the electroweak symmetry since the first
step is the same as that in our previous work \cite{Oda0}. Taking account of the Coleman-Weinberg mechanism of conformal
symmetry and the Wick rotation, the potential  (\ref{V(phi, H)}) is modified to be the following effective potential at the one-loop
level:
\begin{eqnarray}
V_{eff} (\varphi, H) = \frac{5}{9216 \pi^2} \xi^4 \varphi^4 \left( \log \frac{\varphi^2}{\langle \varphi \rangle^2}
- \frac{1}{2} \right)  - \lambda_{H \phi} (H^\dagger H) \varphi^2  
+ \frac{\lambda_H}{2} (H^\dagger H)^2.
\label{Eff-potential}
\end{eqnarray}
Inserting the minimum $\varphi = \langle \varphi \rangle$ and completing the square, the effective potential reads
\begin{eqnarray}
V_{eff} (\langle \varphi \rangle, H) = \frac{\lambda_H}{2} \left( H^\dagger H - \frac{\lambda_{H \phi}}{\lambda_H} 
\langle \varphi \rangle^2 \right)^2 - \frac{1}{2} \left( \frac{\lambda_{H \phi}^2}{\lambda_H} 
+ \frac{5}{9216 \pi^2} \xi^4 \right) \langle \varphi \rangle^4.
\label{Eff-potential 2}
\end{eqnarray}
It is obvious that this effective potential has a minimum at $H^\dagger H = \frac{\lambda_{H \phi}}{\lambda_H} 
\langle \varphi \rangle^2$ due to $\lambda_H > 0$ and $\lambda_{H \phi} > 0$.  Taking the unitary gauge 
$H^T = \frac{1}{\sqrt{2}} (0, v + h)$, this fact implies that the square of the vacuum expectation value of 
the Higgs field, $v^2$, and the square of Higgs mass, $m_h^2$, are respectively given by
\begin{eqnarray}
v^2 = \frac{2 \lambda_{H \phi}}{\lambda_H} \langle \varphi \rangle^2,  \quad
m_h^2  = \lambda_H v^2. 
\label{VEV}
\end{eqnarray}
Using the relation $M_{Pl}^2 = \frac{1}{6} \langle \varphi \rangle^2$, which is obtained at the first step of the
symmetry breaking \cite{Oda0}, the magnitude of the coupling constant $\lambda_{H \phi}$ is given by
\begin{eqnarray}
\lambda_{H \phi} = \frac{1}{12} \left( \frac{m_h}{M_{Pl}} \right)^2 \sim {\cal O} (10^{-33}). 
\label{lambda-H-phi}
\end{eqnarray}
This relation makes it clear that in order to account for the big hierarchy between the electroweak scale and the Planck scale
one needs to take a very tiny value of the coupling constant $\lambda_{H \phi}$.   

To summarize, through the Wick rotation over a scalar field existing in the gravitational sector, we have clarified why
a potential term of the BSM, which corresponds to the Higgs mass term in the Higgs potential of the SM, possesses
the negative coefficient. This phenomenon provides us with an example that the gravitational physics essentially
defined around the Planck scale gives rise to useful information on the SM around the electroweak scale. 
It appears that the existence of a stable vacuum in quantum gravity is relevant to that in the SM.

\begin{flushleft}
{\bf Acknowledgements}
\end{flushleft}
We thank G. Ross for useful discussions. This work has been partially done during the stay in the Corfu Summer Institute,
Workshop on the Standard Model and Beyond, and Dipartimento di Fisica e Astronomia "G. Galilei", Universita degli Studi di
Padova. We thank these institutes for kind hospitality. The work was supported by JSPS KAKENHI Grant Number 16K05327.


\end{document}